\documentclass[12pt]{spieman}  
\usepackage{amsmath,amsfonts,amssymb}
\usepackage{graphicx}
\usepackage{setspace}
\usepackage{tocloft}
\usepackage{subcaption}
\usepackage{lineno}

\title{An AOTF based spectro-polarimeter for observing Earth as an Exoplanet}

\author[a,b*]{Bhavesh Jaiswal}
\author[a]{Swapnil Singh}
\author[a]{Anand Jain}
\author[a,c]{K. Sankarasubramanian}
\author[a]{Anuj Nandi}
\affil[a]{Space Astronomy Group, ISITE Campus, U. R. Rao Satellite Centre, Outer Ring Road, Marathahalli, Bangalore, 560037, India.}
\affil[b]{Indian Institute of Science, Bangalore, 560012, India.}
\affil[c]{Indian Institute of Astrophysics, Bangalore, 560034, India.}

\newcommand\aap{Astronomy \& Astrophysics}

\newcommand\ao{Applied Optics}

\cftpagenumbersoff{figure}
\cftpagenumbersoff{table} 
\begin{document} 
\maketitle

\begin{abstract}
Earth is the only known habitable planet and it serves as a testbed to benchmark the observations of temperate and more Earth-like exoplanets. It is required to observe the disc-integrated signatures of Earth for a large range of phase angles, resembling the observations of an exoplanet. In this work, an AOTF (Acousto-Optic Tunable Filter) based experiment is designed to observe the spectro-polarimetric signatures of Earth. The results of spectroscopic and polarimetric laboratory calibration are presented here along with a brief overview of a possible instrument configuration. Based on the results of the spectro-polarimetric calibration, simulations are carried out to optimize the instrument design for the expected signal levels for various observing conditions. The usefulness of an AOTF based spectro-polarimeter is established from this study and it is found that, in the present configuration, the instrument can achieve a polarimetric accuracy of $<0.3$\% for linear polarization for an integration time of 100 ms or larger. The design configuration of the instrument and the planning of conducting such observations from Lunar orbit are discussed.
\end{abstract}


{\noindent \footnotesize\textbf{*}First author email,  \linkable{bhavesh@ursc.gov.in} }

\begin{spacing}{1}   

\section{INTRODUCTION}\label{introduction}
More than 5000 exoplanets have been discovered so far and several have been characterized for their atmospheres \cite{2010ARA&A..48..631S, 2015PASP..127..941C, 2016Natur.529...59S, 2018Galax...6...51L, 2018haex.bookE.100K}. The majority of these planets may not be habitable owing to their orbital configurations and the intrinsic properties of the planet. However, there are several known planets that orbit their stars in the classical habitable zone, where the equilibrium temperatures are cool enough for liquid water to exist. The search for such Earth-like planets is advancing fast with the help of some of the world's most powerful telescopes, both on the ground and in space. The present decade also holds the promise for testing some of the key technologies like stellar coronagraphs \cite{2010exop.book..111T, 2019AJ....157..132L} which are crucial for observing relatively cooler and much fainter habitable zone planets. These observations will allow the study of starlight reflected from the entire day side of the planet at various phase angles, as the planet revolves around the star.

Before understanding the reflected light from other temperate planets, it is crucial to study and retrieve the information contained in the disc-integrated spectrum of Earth. Earth, being the only known planet to host water and life, serves as a unique testbed for such a study. So far, the disc integrated observations of Earth have been limited to a few spacecraft observations which have flown at large inter-planetary distances like \textit{Galileo} \cite{1993Natur.365..715S}, \textit{EPOXI} \cite{2009ApJ...700..915C}, \textit{DSCOVR/EPIC} \cite{2018RemS...10..254Y}, etc. These spacecraft observations are limited in their phase angle coverage and also lack in terms of polarimetric measurements. There have been ground-based observations of \lq Earthshine\rq\space for the same objective in the visible and NIR bands\cite{2003JGRD..108.4710P, 2006ApJ...644..551T, 2012Natur.483...64S}. Earthshine is the reflected light from the day side disc of the Earth which is again reflected by the night side of the moon and captured by the ground based telescopes. The Earthshine measurements can cover a large range of phase angles (usually $\sim$ 40$^\circ$ to 140$^\circ$) as the Moon revolves around the Earth, and can also be equipped with polarization measurements. However, these observations suffer from the depolarization due to the Lunar surface \cite{1957SAnAp...4....3D, 1968JGR....73..649B}. Satellite-based observations of polarization from clouds and aerosols for local regions of Earth have also been carried out from low Earth orbit satellites \cite{2013AMT.....6..991W}. These observations very well demonstrate the importance of polarization measurements in few selected bands, but miss the global view of the planet and broad wavelength coverage. A comprehensive discussion on previous observations of \lq Earth as an Exoplanet\rq\space and their results can be found in Ref. \citenum{2020plas.book..379R}. Further, there has been a lot of theoretical work towards predicting the spectro-polarimetric signatures of Earth-like exoplanets \cite{2008A&A...482..989S,2012A&A...548A..90K,2022arXiv220505669T}. Several of these predictions, lack a confirmation from the direct observations.

Growing interest in future detections of habitable zone planets has led to several proposals for space-based observations of Earth as an exoplanet. For example, projects like \textit{LOUPE}\cite{2012P&SS...74..202K,2021RSPTA.37990577K} and \textit{EarthShine} \cite{2022JATIS...8a4003B} have been proposed to observe Earth as an Exoplanet from the surface of Moon. \textit{LOUPE} is proposed for spectropolarimetry of an unresolved Earth in 400-800 nm band whereas \textit{EarthShine} will perform imaging and spectral measurements in 400 nm-12.5{\textmu}m. In the present work, we discuss the working of an Acousto-Optic Tunable Filter (AOTF) based spectro-polarimeter in a similar observational 
platform as the previously discussed proposals; from a lunar orbiting platform.

The AOTF is the main component of the proposed experiment and it works on the principle of Bragg diffraction using a birefringent crystal such as TeO$_2$, Quartz etc. It filters the incident light in two diffracted beams which are polarized in mutually perpendicular directions \cite{Harris:69, 10.1117/12.7972821, ghatak_thyagarajan_1989, 1994design}. AOTFs can be tuned for wavelength selection using an external Radio Frequency (RF). This ability of wavelength tuning and polarization sensitivity makes them suitable for the application of spectro-polarimetry. Further, AOTFs are based on solid-state devices, which avoids the need for any moving parts. For use in space-based experiments, the AOTF based spectro-polarimeter can be made into a compact light-weighted instrument. Previous spectroscopic \cite{Korablev:18} as well as polarimetric applications \cite{2015P&SS..113..159R, Belyaev:17} of AOTFs are worth noting here. In the current work, based on the laboratory measurements and the radiative transfer simulations, the possibility of observing Earth as an exoplanet with an AOTF based spectro-polarimeter is discussed. 

In this paper, the methodology for conducting the spectro-polarimetric calibration experiments is briefly discussed in Section \ref{methodology}. The experimental setup is discussed in Section \ref{setup}. The calibration and important results from the experiments are presented in Section \ref{results}. The NIR polarization signals of Earth and a possible instrument configuration, incorporating the instrument response obtained from the calibration results, are discussed in Section \ref{simulations}. The need for further characterization of AOTF is highlighted in Section \ref{future_work}. The work has been summarized with a conclusion in Section \ref{conclusion}.

\section{CHARACTERIZING THE AOTF}\label{methodology}
The AOTF consists of an optical crystal, whose one face is bonded to an ultrasonic transducer. When an RF frequency is applied to the transducer, due to the photo-elastic  effect a sinusoidal perturbation of refractive index in the medium is generated. When the phase matching condition is satisfied, the input light beam is diffracted at a given angle \cite{1994design, 2015ExA....39..445A}. The interaction between the light beam and acoustic wave produces two diffracted (+1 and -1) narrowband components, which are polarized in mutually perpendicular directions. There are two types of AOTFs depending on the propagation direction of the acoustic waves: collinear and non-collinear \cite{1994design}. In this work, a non-collinear, dual beam AOTF having TeO$_2$ crystal is used as the dispersive medium. The diffracted wavelength from the crystal was tuned in the $1000-1700$ nm wavelength ($\lambda$) range. In the following two subsections, the methodology for the spectroscopic and polarimetric calibration of AOTF has been discussed. 

\subsection{Spectroscopic Characterization}
AOTFs have been well established as spectrometers and have been successfully flown in various space missions \cite{10.1117/12.2296244, 2015P&SS..113..159R,  Korablev:18}. AOTF-based spectrometers, SPICAM and SPICAV, have been used extensively to study planetary atmospheres \cite{10.1117/12.451998, 2006JGRE..111.9S03K, 2012P&SS...65...38K}. An AOTF-based spectrometer can be preferred over a traditional spectrometer for space applications due to the absence of any moving component, compactness of the overall design and a moderate spectral resolution of few nanometers. Though AOTF based spectrometers also have disadvantages in terms of smaller optical aperture area, large power requirements etc., the benefits offered in the present design in terms of polarimetric output and compactness of the instrument outweigh the limitations. The spectral performance of AOTFs is characterized by (i) Tuning frequency relation ($\lambda$ vs. RF), (ii) Spectral bandpass ($\triangle \lambda$), (iii) Angular aperture and (iv) Diffraction efficiency \cite{1994design,2015ExA....39..445A}. The tuning frequency relation determines the diffracted wavelength corresponding to the frequency applied via the transducer. It is influenced by the acoustic velocity inside the crystal, birefringence of the crystal and the incident angle with respect to the optic axis. The spectral bandpass of the two output beams from the AOTF depends on the wavelength of diffracted beams and the interaction length between acoustic and lightwave within the crystal. This spectral bandpass determines the spectral resolution of the AOTF and has a wavelength dependence \cite{1994design}.

\subsection{AOTF as a polarizer}\label{aotfaspol}
Although AOTFs are known to be good polarizers \cite{10.1117/12.7972821, 10.1117/12.2055150, Belyaev:17}, in order to characterize the AOTF it is assumed to be represented by a partial polarizer. A partial polarizer produces a partially polarized beam when an unpolarized light is incident on it. This partial polarizer is mathematically represented with a 4$\times$4 Mueller matrix (see equation \ref{eq_mmatrix}; Ref. \citenum{2003isp..book.....D}). This matrix describes the linear relationship between the polarization states of the light beam incident on a polarizing optical element and the emerging light beam after passing through the AOTF. The first term, $M_{00}$ is the output intensity corresponding to completely unpolarized input light. According to Stokes-Mueller calculus, the light is represented by a 4$\times$1 Stokes vector \cite{Azzam:16}.

Consider a partial polarizer where the two orthogonal components of the incident electric field vector are affected by two positive constant factors, $k_1$ and $k_2$, respectively. The Mueller Matrix for this partial linear polarizer is then defined as \cite{2003isp..book.....D},
 
\begin{equation}\label{eq_mmatrix}
M= \begin{bmatrix}
M_{00}  & M_{01}    & M_{02}    & M_{03}    \\
M_{10}	& M_{11} 	& M_{12}    & M_{13}    \\
M_{20}	& M_{21}    & M_{22}    & M_{23}    \\
M_{30}  & M_{31}    & M_{32}    & M_{33} 
\end{bmatrix}
=\frac{\alpha}{2} 
\begin{bmatrix}
1 			& \beta c_2 			& \beta s_2 			& 0 		\\
\beta c_2	& c_2^2+\gamma s_2^2 	& (1-\gamma)c_2s_2 		& 0 		\\
\beta s_2	& (1-\gamma)c_2s_2		& s_2^2+\gamma c_2^2 	& 0 		\\
0			& 0						& 0						& \gamma	
\end{bmatrix}
,
\end{equation}

where $\alpha=k_1^2+k_2^2$, $\beta=\frac{(k_1^2-k_2^2)}{\alpha}$, $\gamma=\frac{(2k_1 k_2)}{\alpha}$, $c_2=\mathrm{cos}(2r)$, $s_2=\mathrm{sin}(2r)$ and $r$ is the angle from the reference direction measured with respect to the horizontal direction in a plane perpendicular to the direction of light propagation. In our case, $r$ takes the value of 0$^\mathrm{o}$ and 90$^\mathrm{o}$. Here, $\alpha$ represents the transmission efficiency, $\beta$ is a measure for the polarizing capability of the AOTF and $\gamma$ is a measure of depolarization. A value of $\alpha = \beta = 1$ signifies that the polarizer is an ideal linear polarizer. As a response to an unpolarized input $[1, 0, 0, 0]^T$, the output Stokes vector is given by $p^\prime = \alpha/2[1,\beta c_2,\beta s_2, 0]^T$.

The combination of wavelength tuning and polarized output makes an AOTF-based instrument a good choice for spectro-polarimetric studies. The detailed theory and spectral characteristics such as resolution, angular aperture etc., of a similar non-collinear, single beam AOTF have been studied in Ref. \citenum{2015ExA....39..445A}. In the present work, the polarimetric calibration of the AOTF has been emphasized along with the results from the dual beam spectral calibration.

\section{EXPERIMENTAL TEST SETUP}\label{setup}
\begin{figure}[h!]
	\begin{center}
		\includegraphics[scale=0.75]{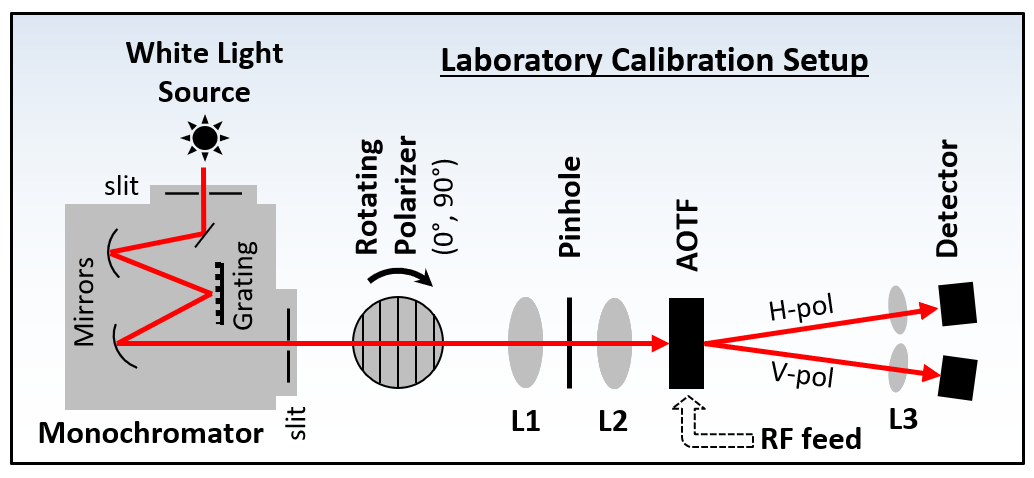}
		\caption{Schematic diagram of the laboratory setup. A broadband light source (halogen lamp) is used along with a grating-based monochromator to produce a monochromatic input beam. Light enters the AOTF after passing through a linear polarizer and a collimator arrangement. The output light from the AOTF consists of H-pol (horizontally polarized e-beam) and V-pol (vertically polarized o-beam) components in two directions. Light is then focused onto an InGaAs detector.} \label{fig_labsetup}
	\end{center}
\end{figure}

In order to characterise the AOTF for its spectro-polarimetric properties, a dual beam AOTF with a 5mm$\times$5mm aperture was used. This AOTF is driven by an external RF driver and the input frequency and power can be controlled via a computer interface. A halogen lamp is used as a broadband light source. A grating-based monochromator is used to create a narrow wavelength output (much narrower than the spectral resolution of AOTF). After the monochromator, an NIR linear polarizer, which has a high extinction ratio of 100,000:1, was used. After the polarizer, the light is collimated using a set of lenses and a pinhole. The diameter of the collimated beam is 3 mm and it fits well within the AOTF aperture. The exit beam from the AOTF is focused onto an Indium-Gallium-Arsenide (InGaAs) detector. The detector output is read via a computer interface. The entire experiment is conducted in a light-proof dark room to minimize background contribution to the measurements. The schematic representation of the test setup is presented in Figure \ref{fig_labsetup} and the major specifications of each of the components used are summarized in Table \ref{tab_setup}.

\begin{table}[h!]
	\caption{Major specifications the elements used in the lab setup.}\label{tab_setup}
	\begin{center}
	\begin{tabular}{ l  l  l   }
		\hline
		 Monochromator  & Resolution            & $\sim$ 0.3 nm \\\hline 
		 Polarizer      & Extinction Ratio      &  100,000:1          \\\hline   
		 Lens L1        & Focal length          & 45 mm     \\
		 Lens L2        & Focal length          & 19 mm     \\
		 Lens L3        & Focal length          & 19 mm     \\\hline
		 Pinhole        & Diameter              & 250 \textmu m \\\hline
		 AOTF           & Type                  & Non-collinear, dual beam \\
		                & Crystal               & TeO$_2$ crystal \\               
		                & Aperture              & 5mm$\times$5mm \\\hline
		 Detector       & Type                  & InGaAs   \\
		                & Quantum Efficiency    & 0.7      \\
		                & Gain                  & 10$^8$ $\Omega$   \\
		                & Responsivity          & 0.98 A/W \\
		                & Operating wavelength range    & 0.9-1.7 \textmu m \\\hline 
	\end{tabular}
	\end{center}
\end{table}

The AOTF has two diffracted beams at the output: (i) extraordinary 
(e) or the horizontally polarized (H-pol) beam and (ii) ordinary (o) or the vertically polarized (V-pol) beam. For the diffracted beam of the AOTF (H-pol as well as V-pol) the observations are conducted in the following sequence:
\begin{enumerate}
    \item Set the desired wavelength in the monochromator.
    \item The polarizer is kept in the vertical position and a scan of AOTF is recorded.
    \item The polarizer is kept in the horizontal position and a scan of AOTF is recorded.
\end{enumerate}
The steps (2) and (3) are then repeated for a \lq without-AOTF\rq\space condition where the AOTF is carefully taken out of the optical path and the detector is moved to the central location. This completes the experiment for one wavelength. These steps are repeated for each wavelength. The same experimental test setup as shown in Figure \ref{fig_labsetup} is used for spectroscopic calibration. For spectroscopic calibration, the polarizer is removed from the optical chain and the spectral response of the AOTF is observed by tuning the input RF.

\section{CALIBRATION RESULTS}\label{results}
\subsection{Spectroscopic Calibration}
\begin{figure}[h!]
\begin{center}
\includegraphics[scale=0.75]{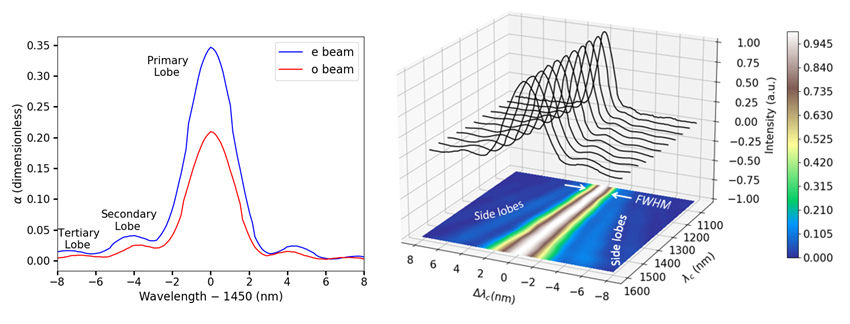}
\caption{Laboratory measurements of the AOTF transfer function ($sinc^2$). \textit{(Left)} Measured transfer function at 1450 nm wavelength for the two beams of AOTF. \textit{(Right)} Spectral variation of AOTF transfer function ($sinc^2$) for e-beam. The x-axis represents the difference from the peak wavelength ($\Delta\lambda_c$) of the $sinc^2$ function and the y-axis represents the wavelength of light input to the AOTF (as central wavelength $\lambda_c$). The projected image in the bottom shows the contours of the measured response. The behaviour of the FWHM and side lobes can be clearly seen across the wavelength.}\label{fig_sinc2}
\end{center}
\end{figure}

Spectroscopic calibration of AOTF mainly involves the study of AOTF transfer function and the frequency tuning relation. In the present design involving two beams of the AOTF, the spectroscopic calibration is done for both beams using the lab setup mentioned in Figure \ref{fig_labsetup} (after removing the polarizer from the path). For the spectroscopic calibration, the intensity profile of the halogen source for a narrow wavelength band (which was selected using the monochromator) was scanned as a function of the RF frequency. The AOTF diffracts light for a narrow wavelength band according to its spectral resolution for that particular wavelength. The shape of the transfer function as well as its peak amplitude (diffraction efficiency $\alpha$) can be different for the two beams. In Figure \ref{fig_sinc2} (left) the transfer function for the \lq e\rq\space and \lq o\rq\space beams are shown for 1450 nm wavelength. Apart from the clear difference in the peak amplitude, there are also minor differences in the shape of the two transfer functions. These differences will ultimately tend to alter the overall transmission of the filter. 

The measurements of the AOTF transfer function were carried out for different wavelengths ranging from 1050 nm to 1600 nm in steps of 50 nm. The intensity profile vs wavelength is shown as the colourmap in Figure \ref{fig_sinc2} (right). The transfer functions for various input wavelengths are shown along the z-axis in Figure \ref{fig_sinc2} and it is observed that the resolution increases with wavelength. The measured intensity profile of the AOTF is similar to the theoretical AOTF transfer function (i.e. $sinc^2$) except for an asymmetry in the side lobes that could be due to crosstalk behaviour or could result from inhomogeneous birefringence of transducer waveguide \cite{1994design, 2006JGRE..111.9S03K, 2015ExA....39..445A}.  

Ref. \citenum{2009OExpr..17.2005M} suggests that this asymmetry could arise when the AOTF crystal is inhomogeneous and light does not travel as a planar wave. They use the sum of an odd number of $sinc^2$ functions (five) to model the AOTF transfer function. The sum of seven or more $sinc^2$ functions was not used due to the limited wavelength range of the scans \cite{2009OExpr..17.2005M} and a large number of free parameters could result in overfitting. For our datasets, we modelled the transfer functions using two models: (I) sum of three $sinc^2$ functions and (II) sum of five $sinc^2$ functions. A comparison at 1250 nm is shown in Figure \ref{fig_tranfunc} of appendix \ref{appA}. We see that the Model (II) fits the secondary and tertiary lobes more accurately when compared to Model (I). Hence for all further analysis, we use Model (II).

\begin{figure}[h!]
\begin{center}
\includegraphics[scale=0.3]{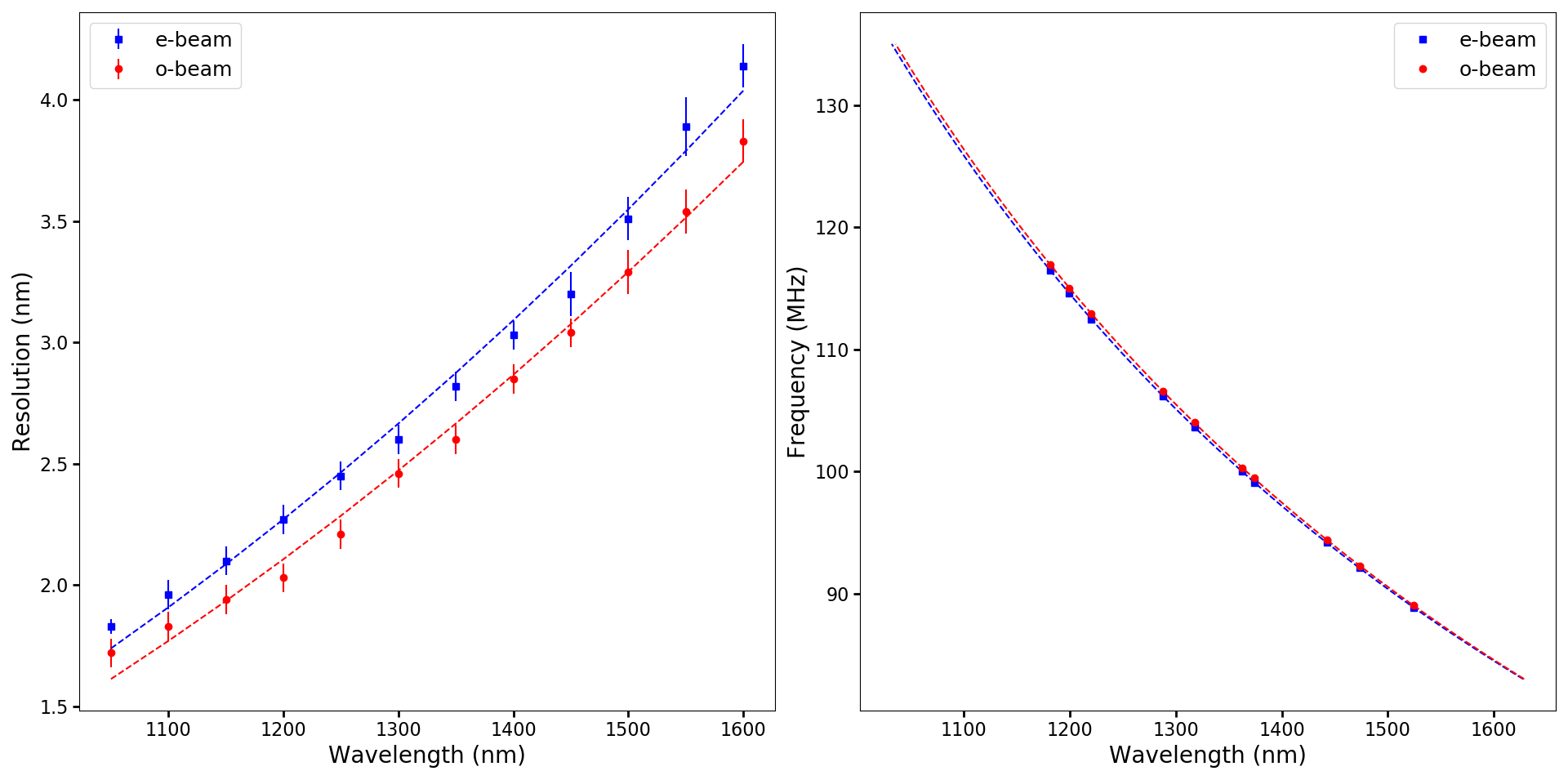}
\caption{\textit{(Left)} Variation of spectral resolution for both AOTF output beams in range of 1.0 to 1.7 $\mu$m. The error bars are 3$\sigma$. The observed variation is fit using the relation between the spectral bandpass and wavelength given in Ref. \citenum{2015ExA....39..445A}. \textit{(Right)} RF-$\lambda$ tuning relation for both e and o beams represented by the dotted lines was obtained by fitting the theoretical tuning relation to the experimentally observed points marked by blue squares for e-beam and red circles for o-beam.}\label{fig_rfl_res}
\end{center}
\end{figure}

To estimate the resolution of the AOTF at various wavelengths, Model (II) function was fitted to the AOTF transfer profiles. It is observed that the resolution of the AOTF ranges from $1.9-4.1$ nm in the wavelength range of $1.0-1.7$ $\mu$m. The variation of resolution with wavelength for both beams of the AOTF is not identical as seen in Figure \ref{fig_rfl_res} (left). This difference between the two beams is a consequence of the difference in the geometry of the acousto-optical interaction for these two beams \cite{Belyaev:17}.

\subsubsection{Frequency Tuning Relation (RF vs $\lambda$)}
In order to obtain the frequency tuning relation with wavelength, the Krypton line source spectrum was acquired with the AOTF. To obtain this spectrum, the monochromator in Figure \ref{fig_labsetup} is replaced with the line source. The spectrum was acquired using the variable RF signal from 85 to 125 MHz and is shown in Figure \ref{fig_lampspectra} of appendix \ref{appB}. The Krypton line source has many lines in the desired wavelength range. Ten Krypton emission lines that do not have any overlap with the neighbouring lines were selected. The frequencies at which these lines are obtained were then matched with the wavelengths of the emission lines to obtain the RF-$\lambda$ calibration by fitting the relation between the tuning frequency and wavelength given in Ref. \citenum{Belyaev:17}. The calibration curves for both beams are shown in Figure \ref{fig_rfl_res} (right). It is observed that the curve for the o-beam closely traces that of the e-beam and the small difference between them could be attributed to the different geometry of the acousto-optical interaction \cite{Belyaev:17}.

\subsection{Polarimetric Calibration}\label{polcal}
\begin{figure}[h!]
\begin{center}
\includegraphics[scale=0.45]{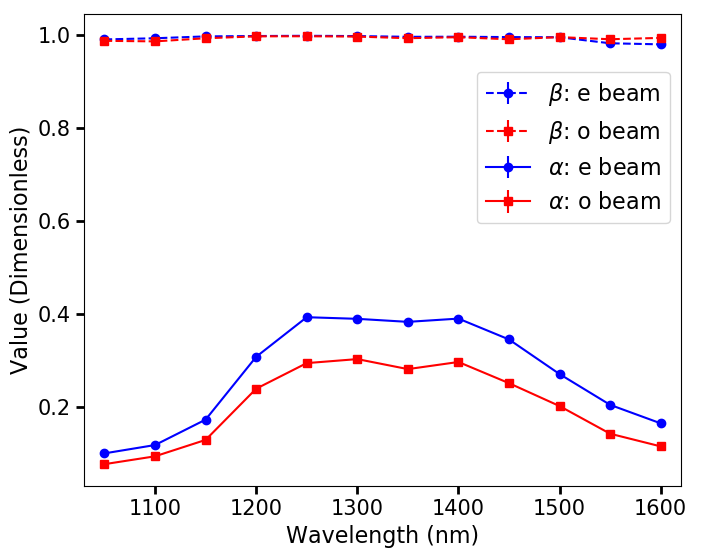}
\caption{Estimated Mueller Matrix parameters for various input wavelengths for both e and o beams is shown in blue squares and red circles respectively. The wavelength dependence of the two major parameters $\alpha$ and $\beta$ which are used to describe the Mueller Matrix of the two beams of the AOTF are represented by solid and dashed lines. The 3$\sigma$ error bars are within the marker sizes.}\label{fig_MMpar}
\end{center}
\end{figure}

The aim of polarimetric calibration is to obtain the Mueller matrix (equation \ref{eq_mmatrix}) of the two beams of the AOTF, across the entire wavelength range. In order to obtain $k_1$ and $k_2$ it is essential to make the transmission measurements of AOTF for the two beams. It is done by making the measurements with and without the AOTF. Without AOTF, signal ($V$) and background ($V^B$) measurements were taken for two positions of the polarizer: (i) $V_{00}$ and $V^B_{00}$ at 0$^\mathrm{o}$ (horizontal) and (ii) $V_{90}$ and $V^B_{90}$ at 90$^\mathrm{o}$ (vertical). With the AOTF, for one beam two datasets ($v$ and $v^{B}$) were taken at both positions of the polarizer: (i) $v_{00}$ and $v^{B}_{00}$ at 0$^\mathrm{o}$ and (ii) $v_{90}$ and $v^{B}_{90}$ at 90$^\mathrm{o}$. The signal comprises of the spectral scan of the central peak of the $sinc^2$ transfer function. The background with the AOTF was measured by fixing the tuning frequency to a value away from the peak frequency value. To remove the contribution of this background, it was subtracted from the signal measurements. Similar datasets were taken for the other beam of the AOTF as well.  

\begin{equation}
    V^\prime_{00}=V_{00} - V^B_{00}\space;\space
    V^\prime_{90}=V_{00} - V^B_{90}
\end{equation}
\begin{equation}
    v^\prime_{00}=v_{00} - v^B_{00}\space;\space
    v^\prime_{90}=v_{00} - v^B_{90}
\end{equation}
Five sets of observations were combined to improve the signal to noise ratio and minimize the error.
Together these datasets provide the values of $k_1$ and $k_2$ for each beam. 
\begin{equation}
    k_1^2=\frac{v^\prime_{00}}{V^\prime_{00}}\space;\space   k_2^2=\frac{v^\prime_{90}}{V^\prime_{90}}
\end{equation}
It is to be noted here that for H-pol beam $k_1$ is obtained by the ratio of H-pol beam (with AOTF) to horizontal beam (without AOTF) whereas $k_2$ is obtained by the ratio of H-pol beam (with AOTF) to vertical beam (without AOTF). And hence $k_2$ is expected to be very close to zero. If $k_2$ = 0, a perfect polarizer with $\beta$ = 1 is obtained. And that is why the values of $k_2$ have to be obtained very carefully as they are very close to the background levels. The measured values of $k_1$ and $k_2$ were then used to estimate the Mueller matrix elements ($\alpha$, $\beta$ and $\gamma$) for both beams at various wavelengths in the 1.0-1.7 $\mu$m range. The errors in $\alpha$, $\beta$ and $\gamma$ have been obtained by propagating the measurement errors in the detector voltages. Their values with errors are listed in \ref{appC}. The variation of $\alpha$ and $\beta$ with wavelength is shown in Figure \ref{fig_MMpar}. 

The variation of $\alpha$ with wavelength for both output beams as seen from Figure \ref{fig_MMpar} suggests that the transmission efficiency is higher for the e-beam when compared to the o-beam. It is also seen that the efficiency is low for lower wavelengths up to 1150 nm, further, it increases and plateaus till 1400 nm and then decreases till 1700 nm. The low efficiency at lower and higher wavelengths is largely due to the variation of the acoustic power, which is incident on the crystal, with the sweep frequency. As the RF is swept from 75 MHz to 150 MHz, the variation seen in the input RF power is $\sim 0.12$ W. The polarimetric efficiency is represented by $\beta$, where $\beta=1$ for an ideal polarizer. From our measurements, $\beta \gtrsim 0.986$ is obtained in the desired wavelength range for both beams and maximum value reaching for $\beta \sim 0.997$. $\beta$ also has information about the linear dichroism properties of the AOTF, which is a measure of the difference in the absorption of linearly polarized light beams with orthogonal planes of polarization. It is observed that the linear dichroism is highest for the wavelength range of 1150 to 1500 nm. These measurements were considered to obtain the signal sensitivity in section \ref{signal_calculations}.

\section{INSTRUMENT RESPONSE TO THE SIGNAL}\label{simulations}
\subsection{Polarized Signal}\label{pol_signal}
Here, we consider a simplistic configuration of an AOTF based spectro-polarimeter with a specific focus on detecting the linearly polarized signal of Earth. The light scattered in planetary atmospheres can get linearly polarized due to Rayleigh or Mie scattering \cite{1974SSRv...16..527H}. The reflected light from planets can get circularly polarized owing to either scattering from clouds \cite{1974SSRv...16..527H} or of biological origin \cite{2009PNAS..106.7816S}. Presently, we ignore the circular polarization as it is usually much smaller than linear polarization. The strength of the scattered linear polarization varies with the phase angle ($\phi$). The phase angle is defined as the star-planet-observer angle as shown in figure \ref{fig_obsgeo}. The Mie scattering linear polarization starts to dominate over that of Rayleigh scattering beyond about 700 nm (for example, see figure 9 of Ref. \citenum{2011A&A...530A..69K}). The maximum polarization occurs at \lq Rainbow angle\rq\space (this is the angle at which Rainbow is formed; for liquid water it is $\sim42^\circ$). This happens due to the process of Total Internal Reflection \cite{2012A&A...548A..90K}, which occurs when light comes out of the cloud drop after one reflection inside the cloud droplets - at the liquid-air boundary. The process of reflection leads to high polarization at these angles and this angle is also indicative of the refractive index of the cloud droplets.

\subsection{Possible Experiment Configuration}\label{experiment}

To observe Earth as a \textit{disc-integrated} planet, the angular extent of Earth needs to be sufficiently small, especially for polarimetric observations because polarization is dependent on the phase angle. When observing Earth from a closer distance, the angular variation of the local regions of Earth disc could be very large, for example it can be about $\sim20^\mathrm{o}$ from a geostationary orbit at $\sim36000$ km altitude. A large variation in the local phase angle can lead to dilution of overall polarization in \textit{disc-integrated} observations.
How much of this local variation can be tolerated can be estimated from the phase angle dependence of the cloud polarization, an example of which is shown in Figure \ref{fig_flux_pol}. Looking at the sharp polarization feature at $42^\mathrm{o}$ it is clear that any disc angle larger than about $\sim10^\mathrm{o}$ would lead to dilution of polarization at this phase angle. Earth, subtending an angle of only $\sim2^\mathrm{o}$ from Moon, makes Moon a choice of preference for such observations.

An ideal experiment to observe Earth as an exoplanet would require to have a broad wavelength coverage (visible to IR) with an ability to do spectroscopic and polarimetric measurements. The space based experiments, however, are often severely constrained in terms of mass and volume of the instrument which also limits the instrument design in many ways. For this reason, we envisage a compact and light weight AOTF based spectro-polarimeter. We chose NIR wavelength band for this experiment as it offers several strong absorption bands\cite{2006ApJ...644..551T} of H$_2$O, CO$_2$, O$_2$ and CH$_4$ which could be of interest to habitability. Measuring the polarization within the absorption bands of different gas species can be of interest as it leads to higher polarization due to reduced multiple scattering \cite{2007A&A...463.1201J,1999JGR...10416843S,2017ApJ...842...41F}. Further, the NIR band has also been shown to be more sensitive to the ocean glints than visible bands\cite{2021A&A...653A..99T,2022arXiv220505669T} though it lacks the sensitivity to Rayleigh scattering. Our choice of NIR band leads us to choose the AOTF and detector which have better performance in this band. We configure the rest of the instrument design with the measured performance of available detectors and AOTF, keeping in mind the compactness of the overall instrument design.
 
\begin{figure}[h!]
	\begin{center}
		\includegraphics[scale=0.75]{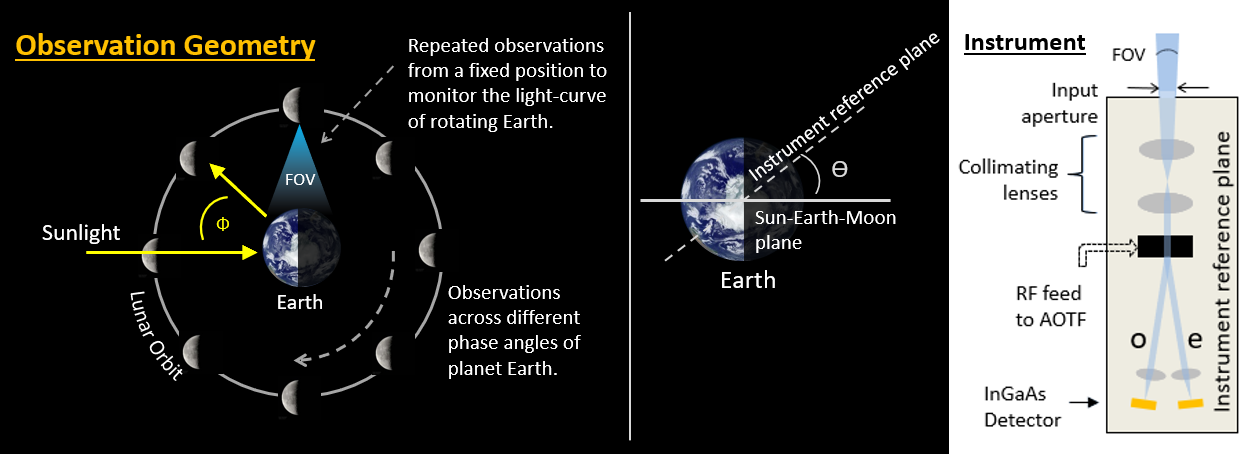}
		\caption{Observation Geometry and instrument configuration. (Left) The orbital configuration of the Moon and the Earth along with the instrument FOV. The relative orientation of the instrument reference plane with respect to the Sun-Earth-Moon plane is also shown. (Right) A functional diagram of the suggested instrument is shown on the right. The input aperture, FOV and instrument reference planes are marked for the instrument. }\label{fig_obsgeo}
	\end{center}
\end{figure}

The suggested experiment consists of a dual beam AOTF based spectro-polarimeter. It would consist of a pair of InGaAs detectors (one for each beam) where the disc of Earth will be focused. The plate scale of the optics can be designed to focus each beam in few pixels. Table \ref{tab_iconf} describes the major specifications of the instrument configuration for the experiment and figure \ref{fig_obsgeo} shows the basic instrument configuration and the observation geometry. As the Moon orbits the Earth, it will be possible to observe different phases of Earth. Similar observations have been carried out using Earthshine measurements \cite{2014A&A...562L...5M, 2019A&A...622A..41S, 2021A&A...653A..99T} but these measurements are susceptible to depolarization from the lunar surface and absorption from the Earth’s atmosphere. Such issues can be eliminated with the direct observations of Earth from the lunar orbit.
 
\begin{table}[h!]
	\caption{Major specifications of the possible instrument configuration for an AOTF-based spectro-polarimetric experiment to observe Earth as an exoplanet.}\label{tab_iconf}
	\begin{center}
	\begin{tabular}{ l  l }
		\hline
		AOTF type           & Non-collinear, dual beam \\
		AOTF crystal        & TeO$_2$  \\\hline
		RF sweep range      & 75-150 MHz \\
		RF step size        & 200 kHz \\
		AOTF input RF power & 1-2 Watts \\\hline
		Detector Type       & InGaAs \\
		Detector QE         &  70\%       \\
		No. of pixels       & 5 \\\hline
		Input aperture      & 2 mm \\\hline
		Spectral range      & $1.0-1.7$ {\textmu}m \\
		Spectral resolution & 2-4 nm \\\hline
		Integration time    & 10 millisec to 1 sec \\\hline
		FOV                 & 2$^\mathrm{o}$ \\\hline
		Light polarization at output & Two orthogonal linear polarizations \\\hline
	\end{tabular}
	\end{center}
\end{table}

\subsection{Radiative Transfer Calculations}\label{RT}
The disc integrated flux and polarization of Earth largely depends upon the extent of the cloud cover and surface features. The cloud cover of Earth is mainly responsible for the polarization features via Mie scattering from cloud droplets whereas land areas usually reflect unpolarized light. A simplistic atmospheric and surface model is considered for the calculations. The model consists of a uniform surface layer covered with gas and clouds. The surface is assumed to be Lambertian having a constant albedo of 0.3 \cite{https://doi.org/10.1002/2014RG000449}. The doubling-adding vector radiative transfer code: PyMieDAP \cite{2018A&A...616A.147R} is used for calculating reflected flux and polarization from Earth's disc. It works by calculating the full Stokes vector of the locally reflected light from the planet. Initially the planet is divided into a grid of $15\times15$ points, where the radiative transfer calculations are performed for each grid point. Next, for the disc-integrated simulations, the locally calculated Stokes vectors are integrated over the parts of the planet which is visible to the observer. The results presented here are for the disc-integrated simulations.

\begin{figure}[h!]
\begin{center}
\includegraphics[scale=0.65]{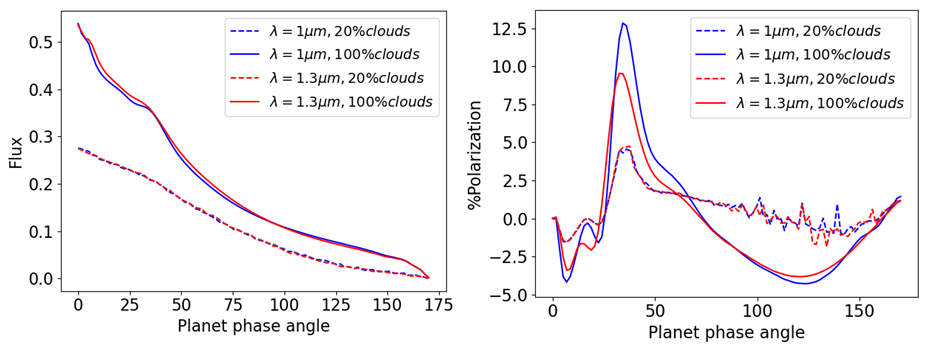}
\caption{Simulation of Flux and Polarization at 1.0 \textmu m and 1.3 \textmu m wavelength. The simulations are carried out at at 100$\%$ and 20$\%$ Cloud cover. (a) Variations of reflected Intensity with phase angles. Total Flux $F$ is given by $\pi\times$ flux for a unit solar flux incident on the planet \cite{2018A&A...616A.147R}. (b) Variations of reflected polarization ($-100Q/I$) with phase angles. The polarization at 1.0 \textmu m for 100$\%$ cloud cover is recreated by simulations from figure 9 of Ref. \citenum{2011A&A...530A..69K}.}\label{fig_flux_pol}
\end{center}
\end{figure}

The strength of polarization as well as spectral features can depend upon various atmospheric features like cloud-top altitude, cloud droplet size, the extent of clouds etc. In this model, we consider standard atmospheric parameters (like pressure and temperature profile) for Earth. The water clouds are kept at an altitude of $\sim$4 km with a cloud droplet size of 6 \textmu m. Patchy clouds\cite{2018A&A...616A.147R} with 20\% and 100\% cloud cover are considered in the simulation to show the extent of polarization variation due to cloud cover (the average cloud cover on Earth is about 50-70\%). The simulation results for flux and polarization are presented in figure \ref{fig_flux_pol} for a fixed continuum polarization wavelength of 1.0 \textmu m and 1.3 \textmu m. The reflected flux is seen to be decreasing, almost linearly, with phase angles (owing to less illumination) and also with decreasing cloud cover (owing to reduced net albedo). The polarization features, as mentioned earlier, show a typical Mie scattering profile where the signal peaks at about 42$^\circ$ for water clouds. Decreasing the cloud fraction uncovers the surface which leads to more unpolarised signal being reflected, leading to less overall polarization. The signals obtained from these simulations are used for further modelling of the instrument response in the next section.

\subsection{Signal Calculations}\label{signal_calculations}
Then disc-integrated signal from Earth at the lunar distance is calculated for each of the two polarised beams of the AOTF and is considered to fall on a total of 5 pixels. The signal calculations were performed using the parameters listed in Table \ref{tab_sigcal}. The total number of electrons generated per pixel in the detector for an integration time $t$ is given by the following relation: 
\begin{equation}\label{eq_signal}
	\mathrm{e}^-/\mathrm{pixel}=(F\times a\times A\times S\times QE\times R\times t)/(\pi\times N\times E_p)
\end{equation}

\begin{table}[h!]
	\caption{Input parameters for signal calculation.}\label{tab_sigcal}
	\begin{center}
	\begin{tabular}{ l  l ll}
	\hline
	Parameter & Symbol & Value & Unit\\\hline
    Solar flux on Earth (@ 1300 nm) & $F$ & 0.3 &W/m$^2$/nm\\
    Telescope aperture diameter  & $a_d$&2& mm\\
    Telescope aperture area & $a$  & $3.141\times10^{-6}$ &m$^2$ \\
    Earth albedo & $A$  & 0.3&\\
    Total no. of pixels on which Earth is focused & $N$  & 5&\\
    Quantum Efficiency of the detector & $QE$  & 0.7&\\
    Earth solid angle from Moon orbit & $S$  & $1\times10^{-3}$ &steradian\\
    Spectral Resolution &$R$  & 3 & nm\\
    Integration time & $t$ & 0.01 to 1 &second\\
    Energy of photon (@ 1300 nm) & $E_p$ & $1.53\times10^{-19}$ & Joule\\\hline
	\end{tabular}
	\end{center}
\end{table}

Considering Mie scattering from spherical particles \cite{1974SSRv...16..527H} the incident signal has only Q polarization if measured in the plane of scattering. Both Stokes U and V are considered zero here. Both Stokes U and V are $\sim$2-3 orders smaller than Stokes Q in visible and IR bands \cite{2020A&A...640A.121G,1974SSRv...16..527H}. Stokes Q however can get converted to Stokes U if the measurements are performed in a plane different from the plane of scattering. If the instrument plane is rotated by angle $\theta$ with respect to scattering plane (Sun-Earth-Moon plane) as shown in Figure \ref{fig_obsgeo}, then the Stokes vector gets multiplied with rotation matrix as follows:
\begin{equation}
	M \times \begin{bmatrix}
		1 & 0 & 0 & 0\\
		0 & \cos2\theta & \sin2\theta & 0\\
		0 & -\sin2\theta & \cos2\theta & 0\\
		0 & 0 & 0 & 1\\
	\end{bmatrix} \times \begin{bmatrix}
		I\\	Q\\	U\\	V\\
	\end{bmatrix}=\begin{bmatrix}
		I_1\\ Q_1\\ U_1\\ V_1
	\end{bmatrix}.
\end{equation}
\newline
Here, I=e$^-/$pixel (see equation \ref{eq_signal}), Q=degree of polarization $*$ I, U and V = 0. $M$ is the Mueller matrix of the AOTF. The detector detects only the first element of the Stokes vector, given by $I_1$. It is noteworthy that the Mueller matrix $M$ can be different for the two beams (as discussed in Section \ref{methodology}), the incident Stokes vector and rotation matrix will remain unchanged though. The intensity detected in the two detectors ($I_1$ and $I_2$) is:
\begin{equation}\label{eq_signal1}
I_1 = \frac{\alpha_1}{2}I + \frac{\alpha_1}{2}\beta_1 Q\cos2\theta
\end{equation}
and
\begin{equation}\label{eq_signal2}
I_2 = \frac{\alpha_2}{2}I - \frac{\alpha_2}{2}\beta_2 Q\cos2\theta .
\end{equation}
Here, the values of $\alpha_1$, $\alpha_2$, $\beta_1$ and $\beta_2$ were obtained for the two beams from Figure \ref{fig_MMpar}. 

\begin{figure}[h!]
	\begin{center}
		\includegraphics[scale=1]{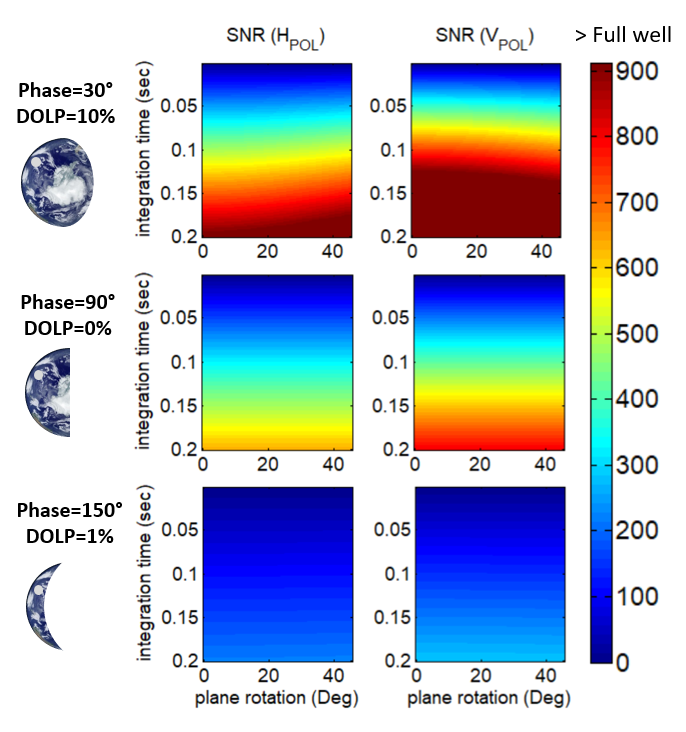}
		\caption{Signal to noise ratio (SNR) simulation for the two detectors (H and V) for various integration times and plane rotation angle. The calculations are shown for different values of incident polarization (P) and phase angles. The detector is saturated at SNR of about 900.}\label{fig_snr}
	\end{center}
\end{figure}

Observed intensity signal in the two detectors is also accompanied by the noise: detector noise and photon noise. Detector noise (including dark noise and read-out circuit noise) is observed to be close to 5000 $e^-/pixel$ for a large range of integration times. In the present electronics circuit, the noise is seen to be dominated by the read-out electronics noise. The photon noise is considered to be $\sqrt{N}$ where $N$ is the no. of photons. The two noises are added in quadrature to obtain the total noise. The calculated signal to noise ratio for the two detectors is shown in Figure \ref{fig_snr} for a range of integration time and plane rotation angle $\theta$. Three different scenarios of polarization signal and phase angles are considered. The SNR mostly depends upon the phase angle of the Earth, the smaller phase angles allow for viewing a large sun-lit portion of Earth and vice versa. For 0$^\mathrm{o}$ phase angle, the instrument views the full day-side Earth disc whereas for 150$^\mathrm{o}$ phase angle only a thin crescent of the day-side Earth is visible to the instrument. The design of the instrument needs a large range of detector integration times which can be tuned for different portions of the orbit in order to maintain the SNR. Also to be noted is the difference in the values of SNR in the two detectors for the same integration time and plane rotation angle is due to the different values of $\alpha$ for the two beams of AOTF (see Figure \ref{fig_MMpar}).

From the observed values of $I_1$ and $I_2$ and by solving equations \ref{eq_signal1} and \ref{eq_signal2}, the estimated desired Degree of Linear Polarization (DOLP) or $Q/I$ of the incident radiation is given as
\begin{equation}\label{eq_DOLP}
\frac{Q}{I} = \frac{\alpha_1 I_2 - \alpha_2 I_1}{\alpha_1 \beta_1 \cos2\theta I_2 + \alpha_2 \beta_2 \cos2\theta I_1} .
\end{equation}

\begin{figure}[h!]
	\begin{center}
		\includegraphics[scale=1]{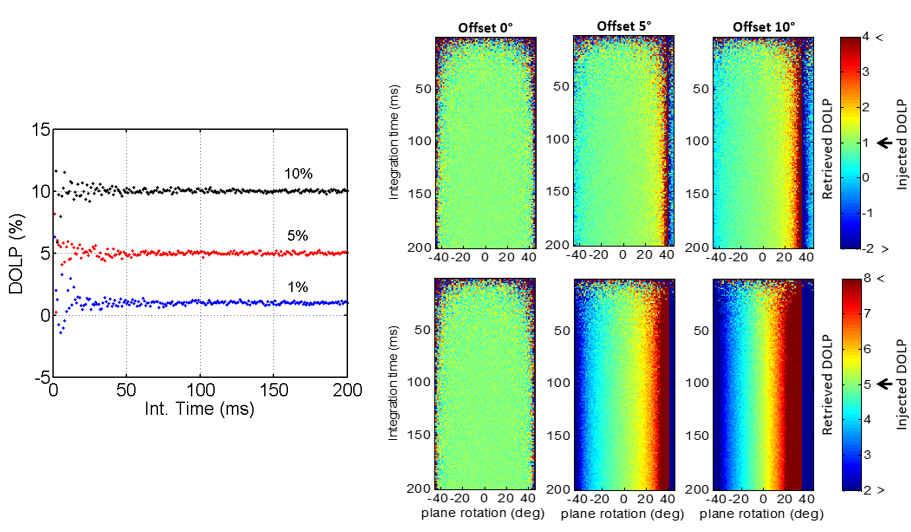}
		\caption{Synthetic retrieval of DOLP from the instrument. \textit{(Left)} Three different values of known DOLP are injected into the instrument model: 1\%, 5\% and 10\% and the retrieved DOLP values are plotted in blue, red and black colors respectively. \textit{(Right)} The effect of platform rotation on the retrieved DOLP for 1\% (top panel) and 5\% (bottom panel) injected DOLP. Three cases are considered for an \lq unknown\rq\space offset of 0$^{\circ}$, 5$^{\circ}$ and 10$^{\circ}$ in the plane rotation angle.}\label{fig_pol}
	\end{center}
\end{figure}

The retrieved DOLP (from equation \ref{eq_DOLP}) is dependent on the measured values of $I_1$ \& $I_2$ which will have errors due to detector noise as well as the photon noise. DOLP is also dependent on values of $\alpha$ and $\beta$ which also have errors due to measurement uncertainties. All these errors combined together lead to a retrieved value which also has uncertainties. A synthetic retrieval of DOLP was performed considering all these sources of uncertainties for a large range of integration times. Three different values of DOLP are injected and corresponding retrieved values for a range of integration times are shown in Figure \ref{fig_pol}. The retrieved values show a distribution centred about the injected value of DOLP. This distribution decreases in width with increasing values of integration time. Large integration times lead to larger SNRs (as in Figure \ref{fig_snr}) and hence leads to better constraints on retrieved values. The measured uncertainties of $\alpha$ and $\beta$ of AOTF are small and have minimum effect on the retrieval. It was found that the noise from the detector is the dominant source of errors. It was also seen that for the present configuration of the instrument, the integration times of about 100 ms and larger can lead to an uncertainty of $\sim$0.3\% in the retrieved DOLP.

The retrieved DOLP also depends upon our knowledge of the angle $\theta$ (via equation \ref{eq_DOLP}). Space based observations can be prone to mis-alignments due to jitter and drift of satellites causing an error in the knowledge of $\theta$. It can underestimate or overestimate the DOLP by a factor of $1/\cos{2\theta}$ (see equation \ref{eq_DOLP}). This can cause a non-linear error in the retrieved DOLP depending upon angle $\theta$. A synthetic retrieval excercise of DOLP is performed by deliberately adding an 'unknown' offset in the angle $\theta$ for various integration times and a range of plane rotation angles ($\theta$ = -45$^{\circ}$ to 45$^{\circ}$). The retrieved DOLP values are shown in the right panel of Figure \ref{fig_pol} for two different cases of injected DOLP (1\% and 5\%). We can see that offset of 0$^{\circ}$ (no offset) allows us to retrieve the true value (injected value) for a large range of plane rotation angles. For an offset of 5$^{\circ}$ we start to see a constant shift in the retrieved values with respect to plane rotation angle. This shift is larger for larger angle offsets and also for larger injected DOLP values. For 10$^{\circ}$ offset the shifts are more amplified. From this analysis we find that error (or shift) in the retrieved DOLP can be within ~15\% of the true value if the satellite platform is maintained within 10$^{\circ}$ about the scattering plane ($\theta$ = 0$^{\circ}$). It is noteworthy that even if the instrument reference plane is tilted at large angles (plane rotation angle), the retrieved polarization is very close to the incident polarization as long as the angles are known.

\section{FUTURE WORK}\label{future_work}

In this work, the usefulness of an AOTF based spectro-polarimeter instrument for observing Earth as an exoplanet has been established. There are more studies that need to be carried out in future for AOTF characterization, especially to understand the temperature related effects. The current experiment was carried out at room temperature but the RF-$\lambda$ tuning relation of AOTF is known to shift with temperature and needs to be calibrated \cite{2006JGRE..111.9S03K, 2012P&SS...65...38K}. Along with that, it may be required to maintain the operating temperature range of AOTF in orbit, so that the calibration does not change significantly. The temperature of AOTF crystal as well as the RF power amplifier are both important for the calibration purposes. The varying temperature of RF power amplifier in space can lead to varying acoustic power which in turn can affect the diffraction efficiency of the crystal \cite{doi:10.1177/0003702819881786}. The study of temperature related effects is planned in the near future and will be taken up as a continuation of current work. The polarimetric calibration, as discussed in this work depends on various other factors such as the lenses, coatings, detector response etc., and hence a final end-to-end calibration of such an instrument will be carried out with flight components before flight. 

For the calculations of the instrument response, the peak value of $\alpha$ (as in Figure \ref{fig_sinc2}) is considered. Since the AOTF transfer function of the two beams is different for the two beams in terms of FWHM (Figure \ref{fig_rfl_res}) and in terms of shape of transfer function (Figure \ref{fig_sinc2}), the effective value of $\alpha$ should take care of these variations, which may alter the transmission of the beams. This can be done by normalizing the two values of $\alpha$ with their areas under the curve. Although, the instrument design uses two identical detectors for the two beams, one may need to correct for the difference in the response of the detector, if any, in a similar manner. The response of each pixel in the detector will need a relative calibration by doing a flat-fielding of the detector. A final calibration of the instrument with an unpolarized source will reveal any relative difference in the transmission of the two channels ('e' and 'o' beams) which may arise due to unknown sources, such as detector response, transmissivity of lenses etc. It may be required to study the stability of such calibration in orbit by regularly observing a known source.

The above mentioned methodology for retrieving polarization, works well for the continuum part of the spectrum. For retrieving the polarization within the absorption bands one may need to convolve the modeled spectrum with the AOTF transfer function. This may be required depending upon the mismatch in the 'shape' of AOTF transfer function of the two beams, as the absorption lines in the spectrum are much narrower than the width of the transfer function. 

\section{DISCUSSION AND CONCLUSION}\label{conclusion}
Spectro-polarimetry of Earth will allow us in bench-marking the spectral and polarimetric signatures of temperate exoplanets against Earth. A possible configuration of an AOTF based spectro-polarimeter experiment is presented in this work. AOTFs have advantages over other wavelength dispersive/polarization measurement systems such as they are compact, devoid of any moving parts, and have the ability to carry out rapid scans with good spectral resolution (2-3 nm). Hence, AOTFs have been widely used in planetary studies and astronomy \cite{2014SPIE.9147E..2TM, 2015AcPPA.127...81Y, 2018ApOpt..57C.103K, 2021Mate...14.3454L}. Here, we have proposed an experiment which serves as a compact instrument for studying Earth as an exoplanet in reflected light over a broad spectral range in two orthogonal polarization directions. The experiment is designed for observations from the Lunar orbit which allows capturing all the phase angles of Earth, mimicking the future observations of directly imaged exoplanets which could also be sampled for a large range of phase angles. 

Considering the uncertainties in measurements, photon noise and detector noise, it is found that the instrument can achieve a polarimetric accuracy of less than 0.3\% for integration times of 100 ms and larger. This is also consistent with previous AOTF based polarimeter observations of Venus clouds \cite{2015P&SS..113..159R} where the errors are consistently close to $\sim$0.1\%. In this design, the major source of uncertainty originates from the detector noise. The values of $\beta$ are very close to 1 with a measurement uncertainty of $<0.1$\% which shows that AOTF itself can achieve much better accuracy if other noises (like photon noise and detector noise) are not significant. The instrument signal to noise ratio can be further improved by co-adding several frames of a single observation. With this methodology and the considerations outlined in the discussion, we propose this experiment as a piggy-back instrument on a future lunar mission to observe the spectro-polarimetric signatures of disc-integrated Earth.

\appendix    

\section{AOTF TRANSFER FUNCTION MODELLING}\label{appA}

\begin{figure}[h!]
\begin{center}
\includegraphics[scale=0.8]{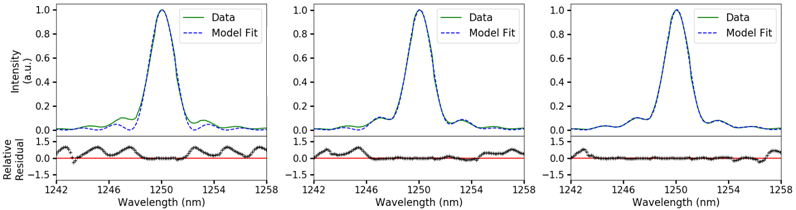}
\caption{AOTF transfer function at 1250 nm modelled using $sinc^2$ function (left), sum of three $sinc^2$ functions (middle) and sum of five $sinc^2$ functions (right). In each plot, the data is shown in green, model is shown with dashed blue lines and the relative residuals  ([Data$-$Model]/Data) are shown in the lower panel. The zero line in the bottom panels is marked using red solid lines.}\label{fig_tranfunc}
\end{center}
\end{figure}

A sum of $sinc^2$ functions were used to model the AOTF transfer function as suggested by \cite{2009OExpr..17.2005M}. Models (I) sum of three $sinc^2$ functions and (II) sum of five $sinc^2$ functions were used to model the transfer functions. From Figure \ref{fig_tranfunc}, it is seen that for the secondary and tertiary lobes Model (II) shows an improvement over Model (I). 

\section{KRYPTON LAMP SPECTRUM}\label{appB}
\begin{figure}[h!]
    \begin{center}
        \includegraphics[scale=0.5]{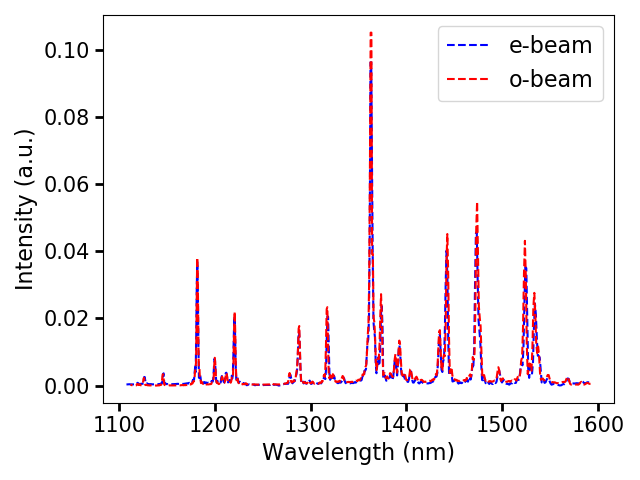}
        \caption{Krypton lamp spectrum obtained for both beams of the AOTF.}\label{fig_lampspectra}
    \end{center}
\end{figure}

A Krypton lamp was used to measure the spectrum for both e- and o-beams in the lab. This spectrum was used to estimate the tuning relation for each of the diffracted beams of the AOTF. It is observed that the spectrum for the e-beam closely traces the spectrum for the o-beam and there is a marginal difference in the tuning relations for the two beams. The measured lab spectra for the two beams is shown in Figure \ref{fig_lampspectra}. 

\section{ESTIMATED MUELLER MATRIX PARAMETERS}\label{appC}
The Mueller Matrix for each of the AOTF beams was estimated using the relation for a partial polarizer as described in Section \ref{aotfaspol}. The Mueller Matrix elements are computed from three major parameters $\alpha$, $\beta$ and $\gamma$. In this work, the variation of these parameters with wavelength was studied (see Section \ref{polcal}). The values of these parameters with wavelength are listed in the following Table \ref{tab_mmpar}. 
\begin{table}[h!]
\begin{center}
\caption{Variation of the Mueller Matrix parameters $\alpha$, $\beta$ and $\gamma$ with wavelength.} \label{tab_mmpar}
\footnotesize
\begin{tabular}{|c|c|c|c|}
\hline
$\lambda$ (nm) 	 & $\alpha_e$ 	 & $\beta_e$ 	 & $\gamma_e$ 	 \\\hline
1050 	 & $ 0.1007 \pm 0.0001 $ 	  & $ 0.9899 \pm 0.0019 $ 	  & $ 0.1420 \pm 0.0084 $ 	 \\
1100 	 & $ 0.1186 \pm 0.0001 $ 	  & $ 0.9924 \pm 0.0014 $ 	  & $ 0.1230 \pm 0.0096 $ 	 \\
1150 	 & $ 0.1732 \pm 0.0001 $ 	  & $ 0.9964 \pm 0.0007 $ 	  & $ 0.0852 \pm 0.0079 $ 	 \\
1200 	 & $ 0.3081 \pm 0.0001 $ 	  & $ 0.9972 \pm 0.0005 $ 	  & $ 0.0748 \pm 0.0044 $ 	 \\
1250 	 & $ 0.3932 \pm 0.0001 $ 	  & $ 0.9975 \pm 0.0004 $ 	  & $ 0.0703 \pm 0.0042 $ 	 \\
1300 	 & $ 0.3897 \pm 0.0001 $ 	  & $ 0.9971 \pm 0.0004 $ 	  & $ 0.0767 \pm 0.0045 $ 	 \\
1350 	 & $ 0.3832 \pm 0.0001 $ 	  & $ 0.9955 \pm 0.0004 $ 	  & $ 0.0947 \pm 0.0036 $ 	 \\
1400 	 & $ 0.3902 \pm 0.0001 $ 	  & $ 0.9957 \pm 0.0004 $ 	  & $ 0.0924 \pm 0.0036 $ 	 \\
1450 	 & $ 0.3454 \pm 0.0001 $ 	  & $ 0.9949 \pm 0.0005 $ 	  & $ 0.1013 \pm 0.0042 $ 	 \\
1500 	 & $ 0.2710 \pm 0.0001 $ 	  & $ 0.9946 \pm 0.0007 $ 	  & $ 0.1037 \pm 0.0072 $ 	 \\
1550 	 & $ 0.2049 \pm 0.0002 $ 	  & $ 0.9814 \pm 0.0012 $ 	  & $ 0.1918 \pm 0.0073 $ 	 \\
1600 	 & $ 0.1652 \pm 0.0002 $ 	  & $ 0.9794 \pm 0.0016 $ 	  & $ 0.2018 \pm 0.0093 $ 	 \\\hline
$\lambda$ (nm) 	 & $\alpha_o$ 	 & $\beta_o$ 	 & $\gamma_o$ 	 \\\hline
1050 	 & $ 0.0774 \pm 0.0002 $ 	 & $ 0.9880 \pm 0.0037 $ 	 & $ 0.1546 \pm 0.0329 $ 	 \\
1100 	 & $ 0.0944 \pm 0.0001 $ 	 & $ 0.9855 \pm 0.0013 $ 	 & $ 0.1695 \pm 0.0087 $ 	 \\
1150 	 & $ 0.1296 \pm 0.0001 $ 	 & $ 0.9924 \pm 0.0008 $ 	 & $ 0.1234 \pm 0.0052 $ 	 \\
1200 	 & $ 0.2396 \pm 0.0001 $ 	 & $ 0.9963 \pm 0.0005 $ 	 & $ 0.0854 \pm 0.0054 $ 	 \\
1250 	 & $ 0.2945 \pm 0.0001 $ 	 & $ 0.9967 \pm 0.0004 $ 	 & $ 0.0815 \pm 0.0044 $ 	 \\
1300 	 & $ 0.3032 \pm 0.0001 $ 	 & $ 0.9960 \pm 0.0004 $ 	 & $ 0.0894 \pm 0.0038 $ 	 \\
1350 	 & $ 0.2817 \pm 0.0001 $ 	 & $ 0.9926 \pm 0.0006 $ 	 & $ 0.1217 \pm 0.0048 $ 	 \\
1400 	 & $ 0.2970 \pm 0.0001 $ 	 & $ 0.9947 \pm 0.0005 $ 	 & $ 0.1024 \pm 0.0032 $ 	 \\
1450 	 & $ 0.2101 \pm 0.0002 $ 	 & $ 0.9900 \pm 0.0012 $ 	 & $ 0.1408 \pm 0.0053 $ 	 \\
1500 	 & $ 0.1733 \pm 0.0001 $ 	 & $ 0.9909 \pm 0.0012 $ 	 & $ 0.1346 \pm 0.0068 $ 	 \\
1550 	 & $ 0.1361 \pm 0.0002 $ 	 & $ 0.9876 \pm 0.0021 $ 	 & $ 0.1567 \pm 0.0097 $ 	 \\
1600 	 & $ 0.1157 \pm 0.0001 $ 	 & $ 0.9932 \pm 0.0018 $ 	 & $ 0.1163 \pm 0.0132 $ 	 \\\hline
\end{tabular}
\end{center}
\end{table}


\subsection{Acknowledgments}
We acknowledge the encouragement and support of DD, PDMSA, URSC and Director, URSC regarding this work. We thank Dr T K Alex for his encouragement. We acknowledge
the initial guidance and support of Anurag Tyagi and Dr Manju Sudhakar towards this work. BJ acknowledges timely help and guidance of Lo{\"\i}c Rossi for the use of PyMieDAP package.

\bibliography{bibliography}   
\bibliographystyle{spiejour}   


\vspace{2ex}\noindent\textbf{Bhavesh Jaiswal} is a scientist at Space Astronomy Group, U. R. Rao Satellite Centre (ISRO), India. He graduated from the Indian Institute of Space Science and Technology (IIST), India, in 2011 with a degree in physical sciences and is currently working towards his PhD at the Indian Institute of Science (IISc), India. His research interest includes developing instrument concepts for studying planets.

\vspace{2ex}\noindent\textbf{Swapnil Singh} is a Scientist at the Space Astronomy Group (SAG), U. R. Rao Satellite Centre (ISRO), India. She is currently working on the development, testing and calibration of instruments for ISRO's planetary missions. She completed her Master's degree in Astronomy \& Astrophysics in the year 2019 from the Indian Institute of Space Science and Technology (IIST), India and carried out her research in the field of extragalactic astronomy. 

\vspace{2ex}\noindent\textbf{Anand Jain} is a scientist at Space Astronomy Group, U. R. Rao Satellite Centre (ISRO), India. He graduated from the Indian Institute of Space Science and Technology (IIST), India, in 2011 with a degree in physical sciences. His area of interest is in the development of astronomical instrumentation. 

\vspace{2ex}\noindent\textbf{Dr. K. Sankarasubramanian} is the Group Head of the Space Astronomy Group, U R Rao Satellite Centre (ISRO), India. He has more than 20 years of post doctoral experience in the field of Solar Physics, Optical Instrumentation including spectro-polarimetry for Solar and other astronomical objects. He is PI of an X-ray spectrometer to be flown in one of the future Indian solar observatory class mission.

\vspace{2ex}\noindent\textbf{Dr. Anuj Nandi} is the Division Head of the Space Science Division of Space Astronomy Group, U. R. Rao Satellite Centre (ISRO), India. He did his PhD at S. N. Bose National Centre for Basic Sciences (SNBNCBS), India. His major research interest includes space-based instrumentation and understanding of physics around compact objects.

\vspace{1ex}

\listoffigures
\listoftables

\end{spacing}
\end{document}